\documentclass[letterpaper, reqno]{amsart} 
\usepackage{graphicx} 
\usepackage{oldlfont}
\usepackage{citesort}
\newtheorem*{ack}{Acknowledgment}
\begin{document} 
\title{Observational Limits on Quantum Geometry Effects} 
\author{Tomasz J. Konopka and Seth A. Major} 
\date{January 2002} 
\address{Department of Physics\\ 
Hamilton College\\ 
Clinton NY 13323 USA} 
\email{smajor@hamilton.edu} 

\begin{abstract} 
Using a form of modified dispersion relations derived in the context
of quantum geometry, we investigate limits set by current observations
on potential corrections to Lorentz invariance.  We use a
phenomological model in which there are separate parameters for
photons, leptons, and hadrons.  Constraints on these parameters are
derived using thresholds for the processes of photon stability, photon
absorption, vacuum \v{C}erenkov radiation, pion stability, and the GZK
cutoff.  Although the allowed region in parameter space is tightly
constrained, non-vanishing corrections to Lorentz symmetry due to
quantum geometry are consistent with current astrophysical
observations.
\end{abstract} 
\maketitle

\section{Introduction}

The quantum description of gravitation is arguably the largest gap in
our understanding of fundamental physics.  In the last decade, a
number of lines of research have offered new insights into the theory
which will supersede quantum theory and general relativity.  For
instance, several approaches predict that space is fundamentally
discrete.  In one approach, eigenvalues of geometric observables have
discrete spectra (see e.g. \cite{lqgrev} for a review).  While it may
not be surprising that the quantization of curved space yields
quantized geometry, it is surprising that present day astronomical
observations {\em already} limit the extent of quantum geometry
effects
\cite{kostelecky,AC,colglash,kifune,kluzniak,bert,aloisio,EMN99,EMN,ACP,steckglash,LJM,AP}.

These quantum geometry effects arise from an imprint of discrete 
underlying space on propagating modes.  Particles with ultra-high 
energies interact with structure on the smallest possible scales 
resulting in corrections to Lorentz symmetry.  Observations on the TeV 
scale offer an opportunity to test the extent of these quantum geometry 
induced Lorentz symmetry corrections.

Lorentz invariance may be the most tested symmetry in Nature.  Given
the wealth of evidence in its support, it may seem obtuse to suggest
that there may be corrections.  However, there are reasons to believe
that this may not be an exact symmetry not the least of which is the
fact that any test of Lorentz invariance (necessarily at finite
energy) leaves an infinite parameter space untested.  Due to the
non-compact nature of the Lorentz group, exact Lorentz symmetry is
untestable.  There are other reasons to suspect exact Lorentz
invariance at all energy scales.  For instance, ultraviolet
divergences in quantum field theory point to new physics at high
energies.  Despite these suggestions, only recently has work in
quantum gravity yielded concrete proposals on how the symmetry might
be modified \cite{kostelecky,AC,GP,AM-TUneut,AM-TUphot,BACKG}.

In this paper we explore the consequences of modified dispersion
relations which are motivated by a study of semiclassical states in
loop quantum gravity by Alfaro, Morales-T\'{e}cotl, and Urrutia
\cite{AM-TUneut,AM-TUphot}.  We study a model in which the usual
dispersion relation of special relativity, $E^{2} = p^{2} + m^{2}$, is
modified by leading order quantum geometry corrections of the form
$\kappa \ell_{p} p^{3}$.  By the semiclassical analysis, the parameter
$\kappa$ is expected to be of order unity.  It determines the
extent of the corrections to Lorentz symmetry while the Planck length,
$\ell_{p}$, sets the scale of the effects.  Due to the sensitivity of
process thresholds, the modifications of particle
dispersion relations are already tested by current astronomical
observations.

What is more spectacular than merely limiting the possible extent of
corrections to Lorentz symmetry is the proposal \cite{acpion} that
this framework may be robust enough to elegantly explain three
incongruities, or paradoxes, between standard model predictions and
observational results: (i) Cosmic rays are expected interact with the
cosmic microwave background (CMB) producing pions and introducing an
upper limit on the observed energy of particles of cosmological
origin.  Known as the GZK cutoff \cite{greisen,ZK} this upper limit
has not been observed.  About twenty events at significantly higher
energies have been reported \cite{bird,takeda}.  Modified dispersion
relations can raise the GZK cutoff \cite{kifune}.  (ii) Ultra high
energy photons of cosmic origin are also expected to interact with
infra red background radiation.  According to some estimates for the
background flux, photons of energy 10 TeV or more should not be seen
due to background induced pair production \cite{PM}.  However, higher
energy events have been reported \cite{tev1,tev2}.  Corrections to
Lorentz invariance provides one explanation for this apparent paradox
\cite{kluzniak}.  (iii) Observations of longitudinal development in
extensive air showers of high energy hadronic particles are apparently
inconsistent with predictions \cite{antonov}.  As proposed in Ref. 
\cite{antonov}, one possible explanation is that high energy neutral
pions become stable.  This may also be explained using modified
dispersion relations \cite{acpion}.

We calculate thresholds for processes involving photons, leptons and
hadrons and, with observational limits, constrain the values for the
$\kappa$ parameters thereby confining the extent of quantum geometry
corrections.  In more detail, in the next section we summarize the
results of Alfaro {\em et.  al.} \cite{AM-TUphot,AM-TUneut}. 
Specifying only general properties of a semiclassical state, such as
flatness above a characteristic scale $L$, the authors find that, in
an analysis of particle propagation, photon \cite{AM-TUphot} and
neutrino \cite{AM-TUneut} dispersion relations are modified.  In
Section \ref{process} we give a brief overview of threshold
calculations before turning to a number of processes including: photon
stability, photon non-absorption, vacuum \v{C}erenkov radiation for
electrons and protons, proton non-absorption, and pion stability.  We
calculate constraints from the threshold calculations to investigate
whether it is possible that observed effects may be accounted for by
Lorentz symmetry corrections.  In \ref{phot} we show that asymmetric
momentum partitioning, first noticed by Liberati, Jacobson, and
Mattingly \cite{LJM}, dramatically affects the constraints on the
dispersion relation modifications.  The results of Section
\ref{process} are summarized in Table \ref{figure_table}.  Finally in
Section \ref{limits}, we apply the constraints together with current
observations to limit the extent of potential Lorentz symmetry
corrections.  We summarize the constraints in the final section and in
Figures \ref{egamma} and \ref{pgamma}.  We find that present day
observations tightly constrain -- but still leave open -- the
possibility of Lorentz symmetry corrections of this form.

Particularly close to the present work is the paper by Jacobson,
Liberati, and Mattingly \cite{LJM} in which many of these results were
summarized.  For the most part the present work agrees with this paper
where the subject overlaps, although this work also includes new
threshold calculations and constraints for proton vacuum \v{C}erenkov
radiation, the GZK threshold, and pion stability.

\section{Modified dispersion relations}
\label{setup}

To determine the action of the Maxwell Hamiltonian operator on quantum
geometry Alfaro {\em et.  al.} specify only very general conditions
for the semiclassical state.  The idea is to work with a class of
states which satisfy the following conditions: (i) The state is
``peaked'' on flat and continuous geometry when probed on length
scales larger than a characteristic scale $L$, $L \gg \ell_{p}$.  (ii)
On length scales larger than the characteristic length the state is
``peaked'' on the classical Maxwell $U(1)$ connection.  (iii) The
expectation values of operators are assumed to be well-defined.  In
addition it is assumed that the geometric corrections to the
expectations values may be expanded in powers of the ratio of the
physical length scales, $L$ and $\ell_{p}$.

States peaked on geometry and the geometric connection are the natural
expectation for semiclassical or coherent states which model flat
space.  The work of Alfaro {\em et.  al.} is a forerunner for the
detailed analysis of semiclassical states.  As the specification of
semiclassical states becomes more precise, and work is well under way
\cite{vz,tt,tw,stw,cr,v,al}, we can expect to check these initial
results.

Expanding the quantum Maxwell Hamiltonian on their states Alfaro
{\em et.  al.} find that for massive neutrinos the particle dispersion
relations are modified with \cite{AM-TUneut} 
\begin{equation}
    \label{neutdisp}
    E^{2}_{\pm} = p^{2} +m^{2} + (\kappa \pm \beta)
    \ell_{p} p^{3}
\end{equation}
where $p$ is the magnitude of the 3-momentum, $\kappa$ and $\beta$ 
are of order unity and $\beta$ is helicity-dependent.  The result is 
computed for a superposition of helicity eigenstates.

For photons the authors find that the quantum geometry effects are 
given, to leading order in $\ell_{p} k$, by \cite{AM-TUphot}
\begin{equation}
    \label{photdisp}
    E_{\gamma}^{2} = k^{2} + 2 \alpha k^{4+2 \Upsilon} \ell_{p}^{2+2
    \Upsilon} \pm 4 \beta \ell_{p} k^{3}
\end{equation}
where $\Upsilon$ parameterizes the scaling of the semiclassical state
expectation value with respect to the gravitational connection.  As
before, the parameters $\alpha$ and $\beta$ are of order unity with
$\beta$ parameterizing helicity dependent corrections.\footnote{The
specification for the class of semiclassical states is more general in
\cite{AM-TUphot} than in \cite{AM-TUneut}.  With the same
specification as the photon case, the neutrino dispersion relation
(\ref{neutdisp}) would have the same form as (\ref{photdisp})
\cite{hugo}.} For the purposes of exploring the leading order
helicity-independent effects in this model of quantum geometry, we
take $\Upsilon = -1/2$ and $\beta=0$.  Helicity effects have been
investigated in Ref.  \cite{GK} and may be treated using similar
methods used in the present work.

To place all the particles we consider on the same footing we take a
phenomenological approach in which the modification to the usual
dispersion is parameterized by a parameter $\kappa_{a}$ for particle
species $a$.  Thus, the effect is simply modeled by the modified
dispersion relation
\begin{equation}
    \label{disp}
    E^{2}_{a} = p^{2}_{a} +m^{2}_{a} + \kappa_{a} \ell_{p} p^{3}_{a}.
\end{equation}
In this equation $p$ is the magnitude of the particle 3-momentum, a
quantity greater than 0.\footnote{Since Eq.  (\ref{disp}) is in terms
of the magnitude, the relation is invariant under time-reversal.  The
modification could also be viewed as a $\kappa \ell_{p} p^{2} E$
term.} For each particle species, the parameter $\kappa_{a}$ is, from
the semiclassical analysis, expected to be of order 1 and can take
either sign.  Of particular interest to determine whether positive
$\kappa_{a}$'s are allowed for massive particles.  In this case,
massive particles would be able to propagate faster than the low
energy speed of light, $c$, which has interesting repercussions for
causality.  Alternately, for either sign of $\kappa_{a}$, the
particles might be able to propagate faster than high energy photons
as long as $\kappa_{a} > \kappa_{\gamma}$.

In terms of bookkeeping, we use $\ell_{p} := \sqrt{4 \pi \hbar
G/c^{3}}$ throughout.  We set the low energy speed of light to one
(``c is c is 1'') and $\hbar$ to one.  The scale of the model becomes
$\ell_{p} \sim 3 \times 10^{-28}$ eV${}^{-1}$.  The effects of the
corrections terms become significant when the correction term is of
the same magnitude as the mass term, i.e. when $p_{crit} \approx
(m^{2}/\ell_{p})^{1/3} \sim 10^{13}, 10^{14},$ and $10^{15}$ eV for
electrons, pions, and protons, respectively.\footnote{Incidentally,
this shows why high precision tests of Lorentz invariance such as in
Refs.  \cite{bear} and \cite{larry} do not offer strong limits on the
$\kappa$ parameters.  The energies are too low.  However, if the
characteristic length scale of the semiclassical state were fixed at
much larger scale - say nuclear - then these high precision tests would
offer limits on the analogous parameters of such a model.}

The dispersion relation of Eq.  (\ref{disp}) is not valid for all
momenta.  Since the class of semiclassical states identified by Alfaro
{\em et.  al.} only reproduces flat geometry on scales larger than
$L$, we can only use the dispersion relation within this
approximation; particles we investigate cannot have wavelengths
shorter than $L$.  Thus, the momenta are restricted by $p_{a} \ll
1/\ell_{p}$.  We remain well within this restriction with $\ell_{p}
p_{a} < 10^{-8}$.

We call the modification of the dispersion relations ``quantum
geometry effects'' since the background quantum geometry only provides
the scale on which new non-linear terms enter the equations of the
motion.  In addition the quantum geometry of the semiclassical state
enjoys no reaction due to propagating modes on the geometry.  It is
clear that beyond the realm of the semiclassical model reaction of the
particle on the geometry must be taken into account and we should
expect new behavior.

Throughout this discussion we assume that the momenta, energies, and
lengths are compared in one inertial frame.  Indeed, if Lorentz
symmetry is inexact then there is presumably a preferred frame
(however see, for instance, Refs.  \cite{KGN,ACnoframe,MS} for
modifications of Lorentz symmetry which do not select a preferred
frame).  For the purposes of this phenomenological study, we work in
the preferred reference frame in which the cosmic microwave background
radiation is isotropic.

Within the context of perturbative quantum field theory it is well
known that modified dispersion relations can yield causality
violations and breakdowns in perturbative quantization (see, for
instance, \cite{KL}).  Since we investigate particles for which
the correction terms play a significant role, these issues are
important.  However, in the following we work in a single inertial
frame and do not have the transformations between inertial frames in
this new context.  So we leave investigations of these issues to
future work.  It would be particularly interesting to precisely
characterize modified dispersion relations which are consistent within
the low energy perturbative framework.

\section{Particle Processes}
\label{process}

On account of the modification in the dispersion relation of Eq. 
(\ref{disp}) quantum geometry effects are potentially observable. 
This is due to the sensitivity of process thresholds to the correction
term for ultra-high energy particles.  We outline the framework of the
threshold calculations then present the details of each of the
processes in turn.

\subsection{Threshold Kinematics}

Suppose we have the 2 particle interaction $a+b \rightarrow c+d$.  We 
assume that energy is conserved, $E_{a}+E_{b}=E_{c}+E_{d}$, and 
3-momentum is conserved, ${\bf p}_{a}+{\bf p}_{b}={\bf p}_{c}+{\bf 
p}_{d}$.  The only unusual bit is in the modified dispersion 
relations.  Although the modified threshold calculations are a 
straightforward application of momentum conservation, energy 
conservation and the modified dispersion relations, the calculations 
are not without surprises.  In fact, even when the outgoing particles 
have the same mass, it may not be energetically favorable to partition 
the momentum symmetrically.  This asymmetry was, to our knowledge, 
first observed by Liberati, Jacobson, and Mattingly \cite{LJM}.  

We almost exclusively use the leading order approximation of Eq.  
(\ref{disp})
\begin{equation}
    \label{ldisp}
    E \approx p+ \frac{m^{2}}{2p}+ \frac{1}{2} \kappa
    \ell_{p} p^{2}.
\end{equation}
It is clear that this approximation applies for high energy particles 
only, when $p \gg m$ so that $m \ll p \ll 1/\ell_{p}$.

The interaction geometry may be strongly affected by the correction
term.  At ultra-high momenta the correction term dominates and, if the
sign of $\kappa$ is negative then the energy of the outgoing particles
may be reduced by including transverse momenta.  However in the
regimes we consider, using 4-momentum conservation and these modified
dispersion relations it is not hard to see that the threshold
interaction geometry is what one would expect: the incoming particles'
momenta are antiparallel and the outgoing particles' momenta are
parallel.  Of course, the momenta must be sub-Planckian.  We also
emphasize that, in these dispersion relations, the $\kappa$-parameters
are of order one.  Since the framework of the semiclassical states
requires that the momenta be far below the Planck scale, we employ
this interaction geometry throughout the remainder of this paper.

We use the modified dispersion relation Eq.  (\ref{ldisp}) in the
conservation of energy.  With energy and momentum conservation, we 
find the kinematic constraint
\begin{equation}
    \label{epcons}
    \frac{m_{a}^{2}}{2 p_{a}} + \frac{1}{2} \kappa_{a}
    \ell_{p} p_{a}^{2} +\frac{m_{b}^{2}}{2 p_{b}} + \frac{1}{2} 
    \kappa_{b}
    \ell_{p} p_{b}^{2}=
    \frac{m_{c}^{2}}{2 p_{c}} + \frac{1}{2} \kappa_{c}
    \ell_{p} p_{c}^{2} +
    \frac{m_{d}^{2}}{2 p_{d}} + \frac{1}{2} \kappa_{d}
    \ell_{p} p_{d}^{2}.
\end{equation}
By momentum conservation $p_{d}=p_{i}-p_{c}$ with
$p_{i}:=p_{a}+p_{b}$ being the available incoming momentum.  Thus the
right hand side of Eq.  (\ref{epcons}) may be expressed as a function
of $p_{c}$ only.

To locate the threshold for one of the incoming particles (typically, 
the other momenta is known), we wish to find the minimum final energy 
as a function of the momentum of the outgoing particle $c$.  This may 
be accomplished by differentiating Eq.  (\ref{epcons}) and finding the 
roots.  In the general two-channel case, this is given by
\begin{equation}
0 = -\frac{m_{c}^{2}}{2 p_{c}^{2}} + \kappa_{c}
    \ell_{p} p_{c} + \frac{m_{d}^{2}}{2 (p_{i}-p_{c})^{2}} + \kappa_{d}
    \ell_{p} (p_{i}-p_{c}) 
\end{equation}
which is a quintic in $p_{c}$.  With the physical root(s) in hand one
can return to Eq.  (\ref{epcons}) to find the threshold momentum. 
However, in the individual processes there are better methods than
attempting to directly solve this quintic.  Perhaps the best way to
see this is in the case of photon stability.  As this case is
straightforward and yet gives asymmetric partitioning we present it in
its entirety.

\subsection{Photon Stability: $\gamma \not\rightarrow e^{+}+e^{-}$}
\label{phot}

In special relativity photon decay is forbidden -- a simple
consequence of the energy and momentum conservation.  However, taking
quantum geometry effects into account it is energetically favorable
for ultra-high energy photons to decay.  Recent observations of
multi-TeV photons \cite{tev1,tev2,crab} provide potential threshold
values for the quantum geometry induced decay.

Applying the threshold framework above, we immediately see that there
are considerable simplifications in Eq.  (\ref{epcons}).  In fact, the 
energy-momentum constraint becomes
\begin{equation}
    \label{phot1}
    \kappa_{\gamma} \ell_{p} p_{\gamma}^{2} = m_{e}^{2}
    \left( \frac{1}{p_{e^{+}}} + \frac{1}{p_{e^{-}}} \right) + 
    \kappa_{e} \ell_{p} (p_{e^{+}}^{2} + p_{e^{-}}^{2}).
\end{equation}
Before differentiating, notice that for negative $\kappa_{\gamma}$, 
the electron $\kappa$-parameter must also be negative.  This opens up 
the possibility that a simple $p_{\gamma}/2$ partition of incoming momenta 
may not be energetically favorable.

This is already clear from the modified dispersion relations for
negative $\kappa_{a}$.  As the momentum of the particle increases the
energy increases.  However, when the momentum is near $p_{crit}$ the
energy increases less rapidly.\footnote{At high momentum, at the limit
of the semiclassical approximation, the energy reaches a maximum. 
Understanding what occurs at these energies requires detailed
knowledge of the dynamics in addition to the semiclassical state of
quantum gravity.} This curve strongly affects the nature of threshold
calculations: At high momentum the slope of the energy steadily
decreases so it becomes energetically favorable to partition the
momentum of the outgoing particles asymmetrically.  By asymmetrically
partitioning momentum one can only slightly increase the energy of the
high momentum particle while drastically reducing the energy of the
lower momentum particle leading to a lower total energy (for more
extensive discussion see Ref.  \cite{tomasz}).

With this observation we re-express the outgoing momenta as
$p_{e^{-}}= p_{o} - \Delta$ and $p_{e^{+}} = p_{o} + \Delta$ with
$p_{o}$ being half the available momentum, $p_{o}:= p_{\gamma}/2$. 
The asymmetry factor $\Delta$ can be no larger than $p_{o}$; $- p_{o}
< \Delta < p_{o}$.  The relation Eq.  (\ref{phot1}) becomes
\begin{equation}
   \label{photEp}
    2 \kappa_{\gamma} \ell_{p} p_{o}^{2} =  \frac{m_{e}^{2} 
    p_{o}}{p_{o}^{2} - \Delta^{2}} + \kappa_{e} 
    \ell_{p}(p_{o}^{2}+\Delta^{2}).
\end{equation}
To find the threshold energy, or minimum energy, for this process
we minimize this equation.  This yields a simple quartic for
$\Delta$ which has two physical roots: $\Delta_{*} = 0$  for the 
symmetric and
\begin{equation}
\Delta_{*}^{2} = p_{o}^{2}\left( 1 - \sqrt{\frac{m_{e}^{2}}{-\kappa_{e} 
\ell_{p} p_{o}^{3}}} \right)
\end{equation}
for the asymmetric partitioning, respectively.  The second root is
only real for negative $\kappa_{e}$.  Since the momenta
asymmetry is bounded by $p_{o}$ there is an upper bound on
$\kappa_{e}$ for this root:
\begin{equation}
    \label{kecond}
\kappa_{e} \leq - \frac{m_{e}^{2}}{\ell_{p} p_{o}^{3}}.
\end{equation}
In addition, when the threshold energy is found using asymmetric root 
another condition is required to ensure that all the momenta are 
non-negative.  This condition is $\kappa_{\gamma} > \kappa_{e}$.

To find the regions of $\kappa$-parameter space where the two roots 
impose constraints, it is worth returning to Eq.  (\ref{photEp}).  As 
observed above, if $\kappa_{\gamma}$ is negative then $\kappa_{e}$ is 
as well.  It is then energetically advantageous to asymmetrically 
partition the momentum, i.e. to use the non-zero root $\Delta_{*}$.  
Thus, when both $\kappa_{\gamma} < 0$ and the inequality of 
(\ref{kecond}) is satisfied the asymmetric configuration is 
applicable.  With the two roots and the energy-momentum constraint, the 
momenta at threshold are
\begin{equation}
    \begin{split}
	p_{\gamma_{*}} &= \left[ \frac{ 8 
	m_{e}^{2}}{(2\kappa_{\gamma}-\kappa_{e})\ell_{p}} \right]^{1/3}
	\text{ for } \kappa_{\gamma} \geq 0  \\
	\text{and} \\
	p_{\gamma_{*}} &= \left[ \frac{- 8 \kappa_{e} 
	m_{e}^{2}}{(\kappa_{\gamma}-\kappa_{e})^{2}\ell_{p}} \right]^{1/3}
	\text{ for } \kappa_{e} < \kappa_{\gamma} < 0  
\end{split}
\end{equation}
where the two thresholds are valid in the regions identified by the 
inequalities.

Given a threshold value $p_{\gamma_{*}}$ for the process, these
equations give relationships between $\kappa_{e}$ and
$\kappa_{\gamma}$.  Since the photon is stable if the incoming energy 
is less than the threshold energy, we can assign inequalities to the 
constraints on $\kappa_{e}$ and $\kappa_{\gamma}$. Thus, the photon 
is stable at the threshold $p_{\gamma_{*}}$ if
\begin{equation}
    \label{photkappas}
    \begin{split}
    &\kappa_{\gamma} - \frac{1}{2} \kappa_{e} - \frac{4
    m_{e}^{2}}{\ell_{p}p_{\gamma_{*}}^{3}} < 0 \text{ for }
    \kappa_{\gamma} \geq 0 \text{ and}\\
    &(\kappa_{\gamma} - \kappa_{e})^{2} + \frac{8 m_{e}^{2}}{\ell_{p}
    p_{\gamma_{*}}} \kappa_{e}<0 \text{ for } \kappa_{e} <
    \kappa_{\gamma} < 0.
\end{split}
\end{equation}
The regions of applicability of the two constraints are inherited from
the momenta thresholds.  These constraints, together with the regions 
of applicability, determine allowed regions in the $\kappa$ parameter 
space. 

\begin{figure}
      \begin{center}
  \includegraphics[scale=.85]{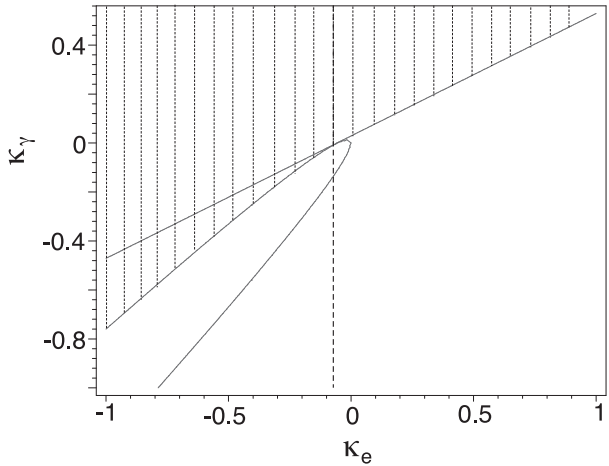}
\end{center}
      \caption{\label{photon_stability} Using the thresholds for 
      photon stability and a threshold value of 50 TeV, the two 
      constraints are plotted.  The two curves smoothly intersect at 
      the transition from symmetric partitioning, the diagonal line, 
      to asymmetric partitioning, the quadratic function in the third 
      quadrant.  The vertical dashed line is the condition of Eq.  
      (\ref{kecond}).  This identifies the upper branch of the 
      asymmetric constraint as the physical solution.  The region 
      ruled out by photon stability up to 50 TeV is indicated with 
      shading.}
\end{figure}

The highest energy observed gamma ray, 50 TeV, originated from the 
relatively local Crab nebula \cite{crab}.  Using this energy for 
$p_{\gamma_{*}}$ we plot the constraints of (\ref{photkappas}) in 
Fig.  (\ref{photon_stability}).  The constraint inequalities ensure 
that the photon does not decay below this threshold.  The shading 
represents regions in $\kappa_{e}-\kappa_{\gamma}$ space which are 
ruled out by the constraints of (\ref{photkappas}).  The smooth 
transition from the linear to the quadratic constraints on the 
$\kappa$ parameters occurs when $\kappa_{\gamma}$ changes sign.  The 
resulting curve forms the upper bound on the allowed region.

There is another process which has a closely related calculation. 
This is the apparent paradoxical observations of TeV photons from 
distant sources.

\subsection{Photon non-absorption: $\gamma + \gamma_{IR}
\not\rightarrow e^{+} + e^{-}$}

Due to the presence of background radiation, high-energy photons are 
expected to be absorbed in pair creation by the far infra-red (IR) 
background.  For photons originating from distances greater than $\sim 
100$ Mpc, the 0.025 eV background provides a cutoff in the spectrum at 
$\sim 10$ TeV. However, recent observations of multi-TeV photons have 
challenged this expectation \cite{tev1,tev2}.  Located at redshift 
$\sim 0.03$ the active galactic nuclei Markarian 421 and Markarian 501 
have produced flares with multi-TeV photons.  Observations have 
included 17 TeV \cite{tev1} and 24 Tev \cite{tev2} events.  It has 
been suggested that the apparent overabundance of high energy photons 
may be due to to corrections to Lorentz invariance along the lines of 
Eq.  (\ref{disp}) \cite{ACP}.

The process kinematics is identical with the case of photon
stability of Eq.  (\ref{photEp}) except the presence of the IR photon. 
Letting the energy of this low-energy photon be $\epsilon$,
conservation of momentum for this process becomes $p_{\gamma} -
\epsilon = p_{e^{+}}+ p_{e^{-}}$ and equation for photon absorption
becomes identical to Eq.  (\ref{photEp}) with the substitution
\cite{LJM}
\[
\kappa_{\gamma} \rightarrow \kappa_{\gamma}' := \kappa_{\gamma} +
\frac{4 \epsilon}{p_{o}^{2} \ell_{p}}.
\]
Thus, the calculation of the threshold is the same as for photon 
stability and resulting $\kappa$ inequalities are identical with the 
substitution $\kappa_{\gamma} \rightarrow \kappa_{\gamma}'$.  Because 
of the factor of $p_{o}$ in the definition of $\kappa_{\gamma}'$, it 
is not as straightforward to write simple expressions for 
$p_{\gamma_{*}}$.  Our main interest is on the allowed regions in 
$\kappa$ parameter space so we are satisfied with the equations which 
relate $\kappa_{e}$ and $\kappa_{\gamma}$ for the allowed region.  
These are presented in Table \ref{figure_table}.  Plots of the allowed 
region are similar in form to the ones for photon stability.

\subsection{Pion Stability: $\pi \not\rightarrow \gamma + \gamma$}
\label{pion}

Within the context of ordinary particle kinematics, the neutral pion 
has a lifetime of $8 \times 10^{-17}$ s.  As suggested by Amelino-Camelia 
\cite{acpion}, the modified dispersion relations can have the effect of 
making ultra-high energy neutral pions stable.  This may explain some 
inconsistencies between observed and expected patterns in cosmic ray 
showers.

Normally the pion decays because the rest mass can be converted into 
energy in the form of photons.  With the modified dispersion 
relations, it is possible that, above a threshold energy, it is no 
longer kinematically possible for the pion to decay.

The threshold calculation is similar to photon stability.  Proceeding 
as before we asymmetrically partition the available momentum for the 
two photons, $p_{o}- \Delta$ and $p_{o}+ \Delta$.  The resulting 
energy-momentum constraint is
\begin{equation}
    \label{pioncons}
    \frac{m_{\pi}^{2}}{2 p_{o}} + 4 \kappa_{\pi} \ell_{p} p_{o}^{2} = 
    2 \kappa_{\gamma} \ell_{p} (p_{o}^{2}+\Delta^{2}).
\end{equation}
As is clear from this equation the analysis splits
into two cases depending on the sign of $\kappa_{\gamma}$.  For
$\kappa_{\gamma}>0$ the minimization requires symmetric partitioning:
$\Delta_{*}=0$.  Solving Eq.  (\ref{pioncons}) for the threshold
momentum we have
\begin{equation}
    \label{pionthresh1}
    p_{\pi_{*}} = \left[ \frac{ 2 m_{\pi}^{2}}{(\kappa_{\gamma}- 2 
    \kappa_{\pi}) \ell_{p}} \right]^{1/3}
    \text{ for } \kappa_{\gamma} > 0 \text{ and } \kappa_{\gamma}> 2 
    \kappa_{\pi}.
\end{equation}
The last inequality ensures that the momentum is positive.

In the second case, $\kappa_{\gamma}<0$, the symmetric partitioning is
in fact the worst choice we could make.  Equal photon momenta is the
maximum of the outgoing energy.  The local minimum for the energy of
the outgoing photons is at the maximum possible asymmetry,
$\Delta_{*}=p_{o}= p_{\pi}/2$, corresponding to the threshold emission
of a zero momentum photon.  From Eq.  (\ref{pioncons}), the
threshold momentum for this process is bounded from below
\begin{equation}
    \label{pionthresh2}
    p_{\pi_{*}} > \left[ \frac{ m_{\pi}^{2}}{(\kappa_{\gamma}-
    \kappa_{\pi}) \ell_{p}} \right]^{1/3} \text{ for }
    \kappa_{\gamma}< 0 \text{ and }
    \kappa_{\gamma}> \kappa_{\pi}.
\end{equation}
The first inequality for $\kappa_{\gamma}$ ensures that the energy
remains positive while the last ensures that the threshold momentum is
positive.

In an identical manner to the previous calculations, Eq. 
(\ref{pionthresh1}) and (\ref{pionthresh2}) yield inequalities among 
$\kappa_{\pi}$ and $\kappa_{\gamma}$.  These are summarized in Table 
\ref{figure_table}. 

\subsection{Vacuum \v{C}erenkov Radiation: $a \rightarrow a + \gamma$}

\v{C}erenkov radiation is observed when a charged particle enters a
medium with a speed greater than the phase velocity of electromagnetic
waves in the medium.  Vacuum \v{C}erenkov radiation in which empty
flat space constitutes the medium is normally forbidden by ordinary
threshold kinematics.  However, Coleman and Glashow suggested that
within the context of a variable-speed-of-light theory, vacuum
\v{C}erenkov radiation could be induced when a charged particle's
speed exceeds the local speed of light \cite{colglash} (see also
\cite{steckglash}).  Vacuum \v{C}erenkov radiation (V\v{C}R) may also
be kinematically allowed by the modified dispersion relations.  This
is clear even from a glance at a charged particle's speed
\begin{equation}
    \label{vel}
    v_{a} = \frac{\partial E_{a}}{\partial p_{a}} = 1 - 
    \frac{m_{a}^{2}}{2 p_{a}^{2}} + \kappa_{a} \ell_{p} p_{a}
\end{equation}
where $a$ is the charged particle.

Since both charged particles and photons have the same functional form 
of the dispersion relations, there are four possible cases to 
investigate.  These depend on the signs of $\kappa_{\gamma}$ and 
$\kappa_{a}$: (i) For $\kappa_{\gamma}>0$ and $\kappa_{a} < 0$ no 
radiation occurs.  (ii) For $\kappa_{\gamma}>0$ and $\kappa_{a} > 0$ 
vacuum \v{C}erenkov radiation may occur when the charged particle's 
speed exceeds 1, the low energy speed of light.  We call this Type I 
V\v{C}R. (iii) For $\kappa_{\gamma}< \kappa_{a}$ vacuum \v{C}erenkov 
radiation occurs either when the charged particle's speed exceeds the 
low energy speed of light or when it exceeds the speed of the 
ultra-high energy photon.  We call this last case Type II V\v{C}R. 
These cases show that V\v{C}R could occur in the first, fourth, and part 
of the third quadrants.

We present the threshold calculation first a charged particle 
$a$.  In Section \ref{limits} we use these results for electrons and 
protons.  The threshold calculations can be found using the methods 
use employed above or, equivalently, using the \v{C}erenkov condition 
$v_{a} > v_{\gamma}$.  Using the later method we find that for high 
energy charged particles the condition for Type I V\v{C}R is
\begin{equation}
    \label{cerenI}
    p_{a_{*}} = \left[ \frac{m_{a}^{2}}{2 \ell_{p} \kappa_{a}} 
    \right]^{1/3} \text{  for } \kappa_{a} > 0.
\end{equation}
This condition corresponds to the momentum at which the particle's 
speed exceeds 1, the low energy speed of light.  This process the 
photon emission begins with long wavelength photons.  For this Type I 
radiation the spectrum extends from zero energy up to 
$p_{\gamma_{max}}=p_{a_{*}}$.

When $\kappa_{\gamma} < \kappa_{a}$, however, the emission process
starts with finite energy photons.  For high energy particles and
photons -- what we call Type II V\v{C}R -- the \v{C}erenkov condition 
yields
\begin{equation}
    \label{cerenthesh1}
    p_{a_{*}} = \left[ \frac{m_{a}^{2}}{2(\kappa_{a} - 
    \kappa_{\gamma}) \ell_{p}} \right]^{1/3}
    \text{ for  } \kappa_{a} > \kappa_{\gamma}.
\end{equation}
This may also be derived using the methods of Section \ref{phot}.  The
observational constraints may be derived from this threshold.

When using the velocity relation, one must also ensure that energy is
conserved.  We study the case in which both outgoing particles have
high momenta.  In the leading order approximation of Eq. 
(\ref{ldisp}), the energy-momentum constraint takes on the familiar
form of the threshold calculations:
\[
\frac{m_{a}^{2}}{p_{a}} + \kappa_{a} \ell_{p} p_{a}^{2} = 
\frac{m_{a}^{2}}{(p_{a}-p_{\gamma})} + \kappa_{a} \ell_{p} 
(p_{a}-p_{\gamma})^{2} + \kappa_{\gamma} \ell_{p} p_{\gamma}^{2}.
\]
After some simplification, this gives a quadratic for $p_{\gamma}$
\begin{equation}
    p_{\gamma}^{2} - \frac{(3 \kappa_{a} + \kappa_{\gamma}) 
    p_{a}}{(\kappa_{a} + \kappa_{\gamma})} p_{\gamma} - 
    \frac{m_{a}^{2}-2 \kappa_{a} \ell_{p} 
    p_{a}^{3}}{(\kappa_{a}+\kappa_{\gamma}) \ell_{p} p_{a}} = 0.
\end{equation}
The requirements that the physical root of this equation,
$p_{\gamma_{*}}$, be real and positive produce the threshold value
\begin{equation}
    \label{cerenthesh2}
    p_{a_{*}} = \left[ \frac{-4 m_{a}^{2}(\kappa_{a} +
    \kappa_{\gamma})}{(\kappa_{a} - \kappa_{\gamma})^{2} \ell_{p}}
    \right]^{1/3} \text{ for } \kappa_{\gamma} < - 3 \kappa_{e} < 0.
\end{equation}
The last inequality arises from the positivity of the maximum energy
outgoing photon
\begin{equation}
    p_{\gamma_{*}}= \frac{(3 \kappa_{a}+ \kappa_{\gamma})}{2 
    (\kappa_{a} + \kappa_{\gamma})} p_{a_{*}}.
\end{equation}

The two types of V\v{C}R have distinct spectra.  For Type I, the
emission of photons starts with very low energy photons and extends up
to the particle's energy $p_{\gamma_{max}} = p_{a}$.  For Type II
V\v{C}R, the spectrum forms a band which begins at the finite energy
corresponding to the condition $v_{a}(p) =c_{\gamma}(p)$ and extends
up to the particle's energy $p_{a}$.  The half angles of the
\v{C}erenkov cone apertures also differ for the two types of
radiation.  The angle may be simply computed from Eq.  (\ref{vel}). 
For Type II,
\begin{equation}
    \theta_{C} \approx \left[ \frac{m_{a}^{2}}{p_{a}^{2}}\left( 1 - 
    \kappa_{\gamma} \ell_{p} p_{\gamma} \right) - 2 \kappa_{a} \ell_{p} 
    p_{a} \right]^{1/2}.
\end{equation}
The angle for Type I is found by setting $\kappa_{\gamma}=0$ in the
above equation.

Using observations of high energy electrons and protons, we can 
establish limits on the the possible values of $\kappa_{e}$ and 
$\kappa_{p}$.  This is done in the final section.

\subsection{Proton Non-Absorption: $p +\gamma \not\rightarrow p+
\pi^{0}$}
\label{gzk}

Shortly after the cosmic microwave background was discovered, Greisen
\cite{greisen} and Zatsepin and Kuzmin \cite{ZK} predicted that the
cosmic ray spectrum should have a cutoff.  They found that cosmic rays
constituents including protons and some heavier nuclei arriving from
cosmological distances would lose energy through interactions with the
CMB photons.  The dominant process, photopion production, provides an
effective energy cutoff for high energy protons from cosmological
sources.  In fact the spectrum was believed to be bounded from above
by $5 \times 10^{19}$ eV. However, the observed spectrum extends well
above this cutoff to $3 \times 10^{20}$ eV. While the number of events
with energies above the GZK cutoff is not large ($\sim 20$), the
locations on the sky are not inconsistent with the hypothesis that the
sources are isotropic.  If these events are indeed particles
originating from cosmological distances then the GZK cutoff must be
modified.  As observed in \cite{kifune}, the modified dispersion
relations considered here can raise the GZK cutoff.

The method for computing the threshold energy for photopion production 
is similar to the above process but the details make the problem 
more complex.  The different masses of the pion and the proton mean 
that much of the simplifications used above do not work.  This returns 
us to the general two-channel case and the resulting quintic for the 
momentum $p_{c}$.  In addition, when $\kappa_{p}$ is negative, large 
asymmetries are favored.  In fact, the threshold process produces 
a low momentum pion which means that we must use the dispersion 
relations of Eq.  (\ref{disp}) instead of the leading order 
approximation.  Hence the constraint on $\kappa_{p}$ was found 
numerically.

Using threshold values in the range $1 \times 10^{19}$ - $8 \times
10^{20}$ eV we performed a search for the value of $\kappa_{p}$ which
separated the parameter space into a region in which the high energy
proton would be stable and a region where it would be unstable.  The
selection criteria were simply that $E_{p} + \epsilon > E_{p}' +
E_{\pi}$, where the energies are expressed using the modified
dispersion relations of Eq.  (\ref{disp}), and $\epsilon$ is the
energy of the cosmic microwave background photon, $7 \times 10^{-4}$
eV. Conservation of 3-momentum was also required.  Since the $\kappa$
parameters for both pion and proton are identical, the procedure
produced a constraint on $\kappa_{p}$ as a function of proton
threshold energy.

This function has the familiar shape of the effective potential of a
particle in a central potential.  At energies below the GZK cutoff
$\kappa_{p}$ is positive, effectively lowering the threshold.  At the
GZK cutoff $\kappa_{p}$ passes through zero.  However, the dependence
of $\kappa_{p}$ on the threshold energy is not a monotonically
decreasing function.  To see this note that since the magnitude of the
proton correction term is larger than the magnitude of the correction
terms of the products, at energies above the GZK cutoff $\kappa_{p}$
must be negative.  At significantly higher energies $\kappa_{p}$
approaches zero from below (as may also be seen with energy-momentum
conservation in the leading order approximation).  The resulting
function has a minimum of $-7.9 \times 10^{-16}$ at $1.6 \times
10^{20}$ eV. Thus, measurements of higher energy cosmic rays will not
restrict the parameter further.  These results are discussed further
in the next section.

\section{Limits on $\kappa$-parameters} 
\label{limits}

With all the above processes, one can derive limits on the 
modifications to the dispersion relations.  The threshold constraints 
of Section \ref{process} determine the allowed regions in 
the $\kappa_{e}-\kappa_{\gamma}$ and $\kappa_{p}-\kappa_{\gamma}$ 
parameter spaces.  These constraints are summarized in Table 
\ref{figure_table}.  With the threshold energies in hand one can plot 
the constraints and determine the allowed regions.  But before 
discussing these plots, we summarize the observational data which 
provide threshold values for each of the processes.

\begin{table}
\begin{tabular}{|l|l|l|}
    \hline \hline
    {\bf Process Type} & {\bf Constraint} & {\bf Applicability} \\
    \hline \hline \hline
    \multicolumn{3}{|l|}{{\em Photon Stability}
    ($\gamma \not\rightarrow e^- + e^+$)} \\ \hline & & \\
    Symmetric momenta &
    $\kappa_\gamma < \frac{1}{2}\kappa_e + \frac{4m_e^2}
    {\ell_p p^3_{\gamma_{*}}}$ & $\kappa_{\gamma} \geq 0$ \\
    Asymmetric momenta &
    $(\kappa_\gamma - \kappa_e)^2
    + \frac{8m_e^2 \kappa_e}{p_{\gamma_{*}}^3 \ell_p} < 0$ &
    $\kappa_e < \kappa_\gamma < 0$ \\
    & & \\
    \hline \hline

    \multicolumn{3}{|l|}{{\em Photon non-absorption}
    ($\gamma + \gamma_{IR} \not\rightarrow e^- + e^+$)}
    \\ \hline & & \\
	Symmetric momenta & $2\kappa_\gamma + \frac{8\epsilon}{\ell_p
	p_{\gamma_{*}}^2}- \kappa_e - \frac{8m_e^2}{\ell_p
	p_{\gamma_{*}}^3} <0 $ & $\kappa_{\gamma} > -\frac{4
	\epsilon}{\ell_{p} p_{\gamma_{*}}}$ \\
    Asymmetric momenta & $ (\kappa_\gamma + \frac{4\epsilon}{\ell_p
    p_{\gamma_{*}}^2} - \kappa_e)^2 + \frac{8m_e^2
    \kappa_e}{p_{\gamma_{*}}^3 \ell_p} < 0$ & $
    \kappa_{\gamma} < -\frac{4 \epsilon}{\ell_{p} p_{\gamma_{*}}}$ \\
    & & \\
    \hline \hline

    \multicolumn{3}{|l|}{{\em Pion Stability}
    ($\pi \not\rightarrow \gamma + \gamma$)}
    \\ \hline & & \\
    Symmetric momenta &
    $\kappa_\gamma < 2\kappa_\pi + \frac{2m_{\pi}^2}{\ell_p 
    p_{\pi_{*}}^3}$ &
    $\kappa_\gamma > 2\kappa_\pi$ and $\kappa_\gamma > 0$ \\
    Asymmetric momenta &
    $\kappa_\gamma < \kappa_\pi + \frac{m_\pi^2}{\ell_p p_{\pi_{*}}^3}$ &
    $\kappa_\gamma > \kappa_\pi$ and $\kappa_\gamma < 0$ \\ & & \\
    \hline \hline

    \multicolumn{3}{|l|}{{\em Proton Stability}
    ($p + \epsilon \not\rightarrow p + \pi$)}
    \\ \hline & & \\
    Numerical Result & $\kappa_p < -8 \times 10^{-16}$ & \\
    & & \\
    \hline \hline

    \multicolumn{3}{|l|}{{\em Vacuum \v{C}erenkov Radiation} ($a
    \not\rightarrow a + \gamma$)} \\ \hline & & \\
    Type I: Zero-energy photon & $\kappa_a < \frac{m_a^2}{2\ell_p 
    p_{a_{*}}^3}$ &
    $\kappa_\gamma > 0$ or $\kappa_e > 0$\\
    Type II: Finite-energy photon &
    $\kappa_\gamma > \kappa_a - \frac{m_a^2}{2\ell_p p_{a_{*}}^{3}}$ &
    $\kappa_\gamma < 0$ \\
    Energy Conservation &
    $(\kappa_a-\kappa_\gamma)^2 + \frac{4m_a^2}{\ell_pp_{a_{*}}^3}
    (\kappa_a+\kappa_\gamma) <0$ &
    $\kappa_\gamma < 0,\kappa_\gamma < \kappa_a,
    \kappa_\gamma < -3\kappa_a$ \\
    & & \\
    \hline \hline

\end{tabular}
\caption{Summary of constraints and their regions of applicability.}
\label{figure_table}
\end{table}

We already discussed in Section \ref{phot} the observation which we 
use for photon stability (the 50 TeV event from the Crab nebula).  We 
now turn to photon non-absorption by the far infra red background.

The unexpected transparency of the universe for high energy photons
may be due Lorentz symmetry corrections (see, however, Ref. 
\cite{jagsteck}).  Although direct measurements from two instruments
onboard COBE and detailed calculations have been completed, the IR
flux is not well-determined \cite{berezinsky}.  This is largely due
to, on the observational side, the flux of IR radiation produced by
interplanetary dust and, on the calculational side, uncertainties in
the galactic-evolution models.  Given the uncertainties in the IR
flux, we only plot an example case.  The photon non-absorption
constraint has a complex dependence on energy.  Nonetheless, in the
relevant region the threshold energy which allows for the maximum
increase in transparency due to suppressed pair creation induced by
background 0.025 eV IR photons is 15 TeV. In the summary plots we use
this energy as a representative case.  If the unexpected transparency
is due to Lorentz symmetry corrections then this constraint is the
lower boundary on the allowed region.

At shorter wavelengths the spectrum is much better established.  In 
fact, with no modifications the universe becomes highly opaque to 
photons above $\sim 100$ TeV from the interactions with the 2.7K cosmic 
background radiation \cite{berezinsky}.  Observations of photons above 
this ``hard limit'' would necessitate more serious consideration of 
photon non-absorption due to modifications to Lorentz symmetry.

As shown in Section \ref{pion} modified dispersion relations can make
the pion stable (see also \cite{colglash,acpion}).  Recently Antonov
{\em et.  al.} reported on a test of Lorentz invariance using
longitudinal development of extensive cosmic air showers
\cite{antonov}.  They found that simulated air showers produced a
better fit to the observed depth of the highest energy air shower
\cite{bird} if the pion became stable at high energies.  This
stability could be achieved if pions are stable above $p_{\pi_{*}}
\sim 10^{18}$ eV. We use this threshold value in the
$\kappa_{p}-\kappa_{\gamma}$ plot.  This becomes a very restrictive
limit.

Vacuum \v{C}erenkov radiation provides a strong restriction ``from the
bottom'' in $\kappa_{a}-\kappa_{\gamma}$ parameter space.  We consider
limits on the allowed region in the parameter space for both electrons
and protons.  The limit for electrons comes from observations of a
supernova remnant.  The best fit for the X-ray spectrum from the
supernova remnant SN 1006 arises from assuming that the radiation is
synchrotron radiation from electrons with energies up to $\sim 100$ TeV
\cite{teve,koyama}.  The highest energy primary cosmic rays are
believed to be protons \cite{antonov}.  As we see next the cosmic ray
spectrum reaches $3 \times 10^{20}$ eV. We use this limit for
V\v{C}R for protons.
 
Finally, cosmic rays above the GZK cutoff are observed.  This cutoff
may be raised by modified dispersion relations.  Using the highest
observed cosmic ray, $3 \times 10^{20}$ eV for the threshold, we would
obtain the constraint $\kappa_{p} < - 5.6 \times 10^{-16}$.  There are
two interesting features to this result.  First, there is minimum at
$1.6 \times 10^{20}$ eV in the $\kappa_{p}$-energy curve discussed in
Section \ref{gzk}.  To prevent a gap in the cosmic ray spectrum
between $1.1 \times 10^{20}$ and $3 \times 10^{20}$ eV we use the
value of $\kappa_{p}$ at the minimum, $-8 \times 10^{-16}$.  Second,
even when one accounts for the uncertainties reported in Ref. 
\cite{bird} the constraint is not consistent with $\kappa_{p}=0$. 
Attributing the raising of the GZK threshold modifications of the
dispersion relations restricts the hadronic parameter to negative
values and effectively rules out exact Lorentz invariance.

\begin{figure}
      \begin{center}
  \begin{tabular*}{\textwidth}{c@{\extracolsep{\fill}}
  c@{\extracolsep{\fill}}}
  \includegraphics[scale=.85]{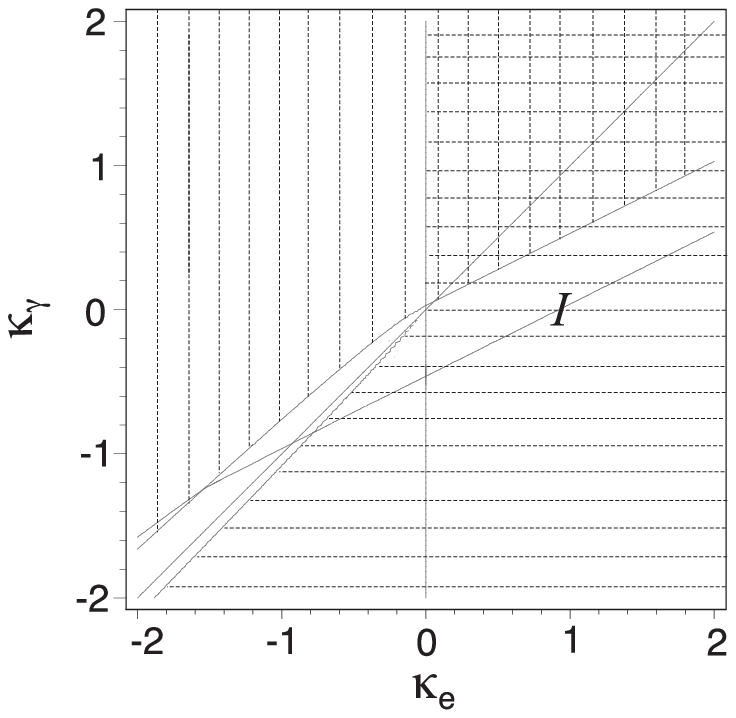}
    &
  \includegraphics[scale=.85]{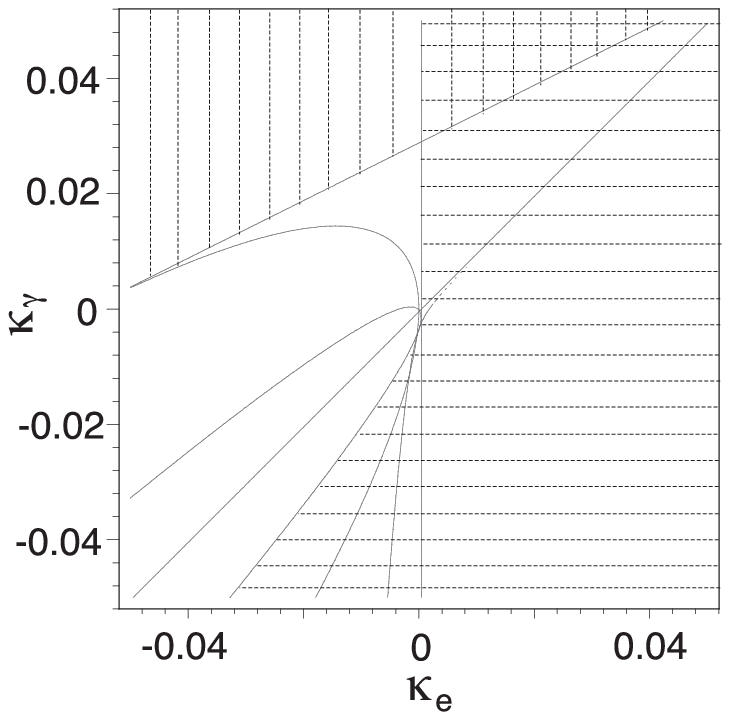}\\
  (a.) &  (b.)
  \end{tabular*}
  \end{center}
\caption{\label{egamma} Plots of the allowed regions in
$\kappa_{\gamma}-\kappa_{e}$ parameter space.  The unshaded portion
shows the region in which corrections to Lorentz invariance are
consistent with current observational limits.  In (a) limits from
photon stability, vacuum \v{C}erenkov radiation, and photon
non-absorption are shown.  The vertical shading corresponds to the
photon stability constraint while the horizontal shading corresponds
to vacuum \v{C}erenkov radiation.  The constraint labeled $I$ arises
from assuming that high energy photons are not absorbed in pair
creation.  If so, this provides a lower limit on the allowed region. 
In (b) the same constraints are shown in a region around the origin. 
The shading is the same as in (a).}
\end{figure}

\begin{figure}
      \begin{center}
  \includegraphics[scale=.85]{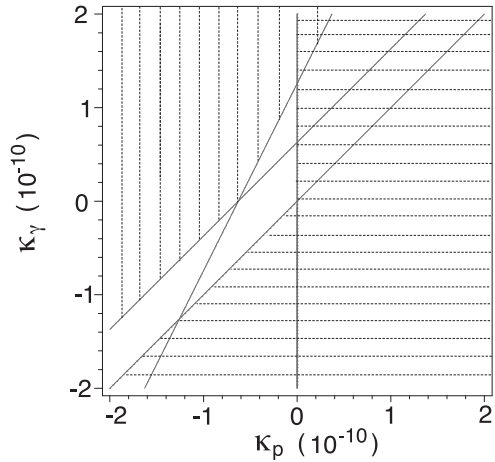}
  \end{center}
\caption{\label{pgamma} The plot of the allowed region in the 
$\kappa_{\gamma}-\kappa_{p}$ space.  Constraints arising from pion 
stability, proton non-absorption, and vacuum \v{C}erenkov radiation 
are shown.  The vertical shading is due to the pion stability 
constraint while the horizontal shading comes from V\v{C}R. The 
constraints due to proton non-absorption and V\v{C}R appear to be a 
single vertical line.}
\end{figure}

In Figures (\ref{egamma}) and (\ref{pgamma}) we identify the allowed
regions of $\kappa_{e}-\kappa_{\gamma}$ and
$\kappa_{p}-\kappa_{\gamma}$ parameter space.  We include all the
above processes, including the more speculative processes of the
photon non-absorption, proton non-absorption, and pion stability. 
This allows us to check whether a consistent set of parameters exist
which simultaneous explain the three paradoxes mentioned in the
introduction and are consistent with current observations.

In the electron-photon plot of Figure \ref{egamma} we include
constraints from photon stability, vacuum \v{C}erenkov radiation, and
photon non-absorption.  The allowed region is a tightly constrained
area mostly lying in the third quadrant and containing the origin. 
Thus, based on these processes it appears that quantum geometry
effects reduce the energy (and speed) of particles at high momentum. 
As Fig.  (\ref{egamma}b) reveals, however, there is a region in which
positive values of the parameters are allowed.  The allowed region
contains the origin and a tiny sliver of positive $\kappa_{e}$
corresponding to superluminal electrons ($\kappa_{e} < 4.5 \times
10^{-4}$).  The $\kappa_{\gamma}$ parameter is not as tightly confined
around the origin and is less than $\sim 0.03$.

In the hadron-photon plot of Figure \ref{pgamma} the limits are
obtained from the processes pion stability, proton non-absorption, and
vacuum \v{C}erenkov radiation.  The quadratic V\v{C}R constraints do
not show up on this scale due to the high upper limit on proton
energy, $3 \times 10^{20}$ eV. In this plot the allowed region is a
exceedingly narrow band again mostly in the third quadrant; the scale
is on the order $10^{-10}$.  There is a small region of positive
values of the parameters allowed.  The band of allowed values is
constrained by the pion stability from above and V\v{C}R from below. 

If these two constraints are borne out by further observations,
different modifications of Lorentz invariance of this form for
hadronic and massless particles would be effectively ruled out.  It is
also interesting that the case $\kappa_{p}=\kappa_{\gamma}$ lies
nearly on the boundary of the allowed region.  The difference of this
relation and the V\v{C}R constraint is $\sim 10^{-15}$. 

We also note that the allowed region does {\em not} contain
$\kappa_{p}=0$, the Lorentz symmetric case.  This is due to the
raising of the GZK threshold.  It is likely that the GZK threshold may
be explained by other means.  Nonetheless, if the GZK cutoff is raised
by corrections of this form, then current observations rule out exact
Lorentz symmetry.  So if we take the three conditions of pion
stability, proton non-absorption, and V\v{C}R seriously, then the
Lorentz symmetry is broken and the parameters are extremely finely
tuned to each other.  From the plot and the constraints on $\kappa$
from the semiclassical analysis, we conclude that $\kappa_{p} =
\kappa_{\gamma} \neq 0$.

Finally, it is also clear that nearly all the positive $\kappa$ space
is ruled out.  In particular, the results for massive particles
effectively rule out ``faster than light'' behavior.  One reason why
this is significant is that satisfies the classical notion of
causality thereby satisfying one of the requirements for consistency
in perturbative quantum field theory.

It is important to keep in mind that these are preliminary results. 
Our strategy is to combine all processes to see whether the results
are consistent.  But it is clear that these constraints are not
equally precise.  For instance the vacuum \v{C}erenkov radiation
constraints are tentative.  The limits for electrons are derived from
indirect measurements of synchrotron radiation around a single
supernova remnant.  We need more data on high energy electrons and
cosmic rays before these processes can truly rule out regions of the
parameter space.

Our V\v{C}R limit for hadrons could also be called ``soft.''  We
assume that the highest energy cosmic ray event was due to a proton. 
As new cosmic ray data becomes available we will be able to better
judge the validity of this assumption.  Once the Pierre Auger project
\cite{auger} is running we should have a much better idea of the value
of the threshold.  Aside from data from more ultra-high energy events,
this detector should also determine the mass composition of the
primary cosmic rays.  This is absolutely critical for the study of
Lorentz symmetry modifications.

Much better estimates of the far infra red background are needed
before we can establish the lower limit of the allowed region in
$\kappa_{e}-\kappa_{\gamma}$ parameter space.  Without this bound we
can not confine the parameters to a finite area.  In Ref.  \cite{LJM}
the uncertainty is the background is handled in a different way. 
Assuming that the actual threshold lies between 10 TeV and 20 TeV they
confine the allowed area to a finite size.

The pion stability constraint will also be improved with new cosmic ray
data.  The results of Ref.  \cite{antonov} compare simulated
showers with the single, highest energy cosmic ray event \cite{bird}. 
It will be very interesting to see whether the onset of pion stability
is observed in other ultra-high energy showers.

The only constraint which could be considered a hard limit comes from
photon stability.  This sharply restricts the possible superluminal
behavior of ultra-high energy photons.  As higher energy gammas are
observed from local sources, this constraint will ``push down'' on the
allowed region.

\section{Conclusions} 
\label{disc}

Beginning with a form of modified dispersion relations, we use exact
energy-momentum conservation, to derive thresholds for particle
processes.  The dispersion relations we consider are motivated by a
class of semiclassical states of loop quantum gravity which are
classical and flat above a characteristic length scale
\cite{AM-TUneut,AM-TUphot}.  We work in a model in which there are
separate parameters for modifications of the dispersion relations for
photons, leptons, and hadrons.  The process thresholds provide a
sensitive test of the quantum geometry effects.  In some cases, such a
photon decay and vacuum \v{C}erenkov radiation, normally forbidden
processes are activated at high energy.  In other cases, such a pion
decay, processes are suppressed.

Current TeV-scale observations offer severe restrictions on the nature
of these effects.  There exist allowed regions in both parameter
spaces consistent with all of the above processes.  The results are
summarized in Figures \ref{egamma} and \ref{pgamma} and in the last
section.  In particular, including all of the processes the analysis
shows that the hadronic parameter is effectively equal to the photon
parameter and both of these are non-zero.  Further the lepton
parameter is confined to a band in the third quadrant as shown in Fig. 
\ref{egamma}.

\begin{ack}
We thank members of the Hamilton College Department of Physics, the
Perimeter Institute, Hugo Morales-T\'{e}cotl, and David Mattingly for
helpful discussions during this work.  T.K. was supported, in part, by
the Ralph E. Hansmann Science Students Support Fund of Hamilton
College.
\end{ack}

\end{document}